\documentclass[10pt]{bmc_article}    
\usepackage{multirow}
\usepackage{float}
\usepackage{graphicx}
\usepackage{amssymb}
\usepackage{commath}
\usepackage{lscape}
\usepackage{color,soul}
\usepackage{url}
\usepackage[figuresright]{rotating}
\usepackage{cite} % Make references as [1-4], not [1,2,3,4]
\usepackage{url}  % Formatting web addresses  

\newenvironment{bmcformat}{\baselineskip20pt\sloppy\setboolean{publ}{false}}{\baselineskip20pt\sloppy}

\newboolean{publ}

\begin{document}
\begin{bmcformat}
	
\title{Using novelty-biased GA to sample diversity in graphs satisfying constraints}

\author{
    Peter Overbury
    \and
    Luc Berthouze\correspondingauthor
    \email{Luc Berthouze\correspondingauthor - L.Berthouze@sussex.ac.uk}
}

\address{
Centre for Computational Neuroscience and Robotics, University of Sussex, UK
}

%\author{
%	Maria Botcharova$^{1,2}$
%	\and
%	Luc Berthouze$^{3,4}$
%	\and
%	Matthew J. Brookes$^5$
%	\and
%	Gareth R. Barnes$^6$
%	\and
%	Simon F. Farmer\correspondingauthor$^2$
%	\email{Simon F. Farmer\correspondingauthor - s.farmer@ucl.ac.uk}
%}

%\address{
%\iid(1)Centre for Mathematics and Physics in the Life Sciences and Experimental Biology/University College London, London, UK \\
%\iid(2)Institute of Neurology/University College London, London, UK \\
%\iid(3)Centre for Computational Neuroscience and Robotics/University of Sussex, Falmer, UK \\
%\iid(4)Institute of Child Health/University College London, London, UK \\
%\iid(5)Sir Peter Mansfield Magnetic Resonance Centre, School of Physics and Astronomy, University of Nottingham, Nottingham, UK\\
%\iid(6)The Wellcome Trust Centre for Neuroimaging, University College London, London, UK
%}

\maketitle

\begin{abstract}

The structure of the network underlying many complex systems, whether artificial or natural, plays a significant role in how these systems operate. As a result, much emphasis has been placed on accurately describing networks using network theoretic metrics. When it comes to generating networks with similar properties, however, the set of available techniques and properties that can be controlled for remains limited. Further, whilst it is becoming clear that some of the metrics currently used to control the generation of such networks are not very prescriptive so that networks could potentially exhibit very different higher-order structure within those constraints, network generating algorithms typically produce fairly contrived networks and lack mechanisms by which to systematically explore the space of network solutions. In this paper, we explore the potential of a multi-objective novelty-biased GA to provide a viable alternative to these algorithms. We believe our results provide the first proof of principle that (i) it is possible to use GAs to generate graphs satisfying set levels of key classical graph theoretic properties and (ii) it is possible to generate diverse solutions within these constraints. The paper is only a preliminary step, however, and we identify key avenues for further development.

\end{abstract}

% How many keywords max? 
%\keywords{Network generation, degree distribution, clustering, higher-order structure, novelty search, coevolutionary dynamic, graph diversity, multi-objective optimisation}

%========================================================================================================
\section{Introduction}
%========================================================================================================
Almost any complex system that involves the interaction of constituent components can be represented as a network. The structural properties of this network will fundamentally affect the way the system operates and thus an accurate description of these properties is often essential to our ability to predict and control the system's behavior. Not surprisingly, there has been much emphasis on applying network theoretic metrics to seek to explain the behaviour of complex systems, e.g., brain connectomics~\cite{Sporns2005}.  When it comes to generating networks with similar properties, however, the situation is very different. There exists only few network generating algorithms, see~\cite{Bansal:2009aa,Newman:2010aa,House2010} and Refs 10, 24, 35, 38, 45 in~\cite{Green2010}, and these are typically only capable of creating networks that satisfy a limited number of classical network characteristics such as global clustering (the propensity of nodes that share a commmon neighbour to be connected), degree distribution (the probability distribution of the number of connections attached to each node) and, more recently, motif distribution for a limited set of such motifs, see Refs in~\cite{Ritchie:2014ab}. In terms of how realistic and usable the generated networks are, these algorithms have two major limitations. First, they are deterministic by construction and are only concerned with fitting the constraints, as opposed to generating diverse networks that fit the constraints (but see~\cite{DelGenio2010} or \cite{Milo2003} for approaches to uniform sampling of graphs with given arbitrary degree sequence). Second, and compounding the above, there is increasing evidence that controlling for classical indicators such as degree distribution and global clustering does little by way of accounting for important structural differences including local clustering, betweenness centrality, higher-order structure, all of which can have significant effects on the dynamics of a network~\cite{Ritchie:2014aa}. There is therefore a need for a more open ended or flexible method for generating networks, one which will not only satisfy a set of criteria but also provide the means to sample the diversity of the space of solutions (it is assumed that, save for the smallest networks, exhaustive mapping of all valid networks is unrealistic). 

As a population-based method of searching the feature space~\cite{Konak:2006aa}, genetic algorithms (GA) are in principle well suited to deal with this multi-objective problem, and indeed, they have been used in a number of related but distinct problems. Risi et al.~\cite{Risi:2012aa} use Hybercube-based NeuroEvolution of Augmenting Topologies (HyperNEAT) to evolve artificial neural networks (ANN) with a particular substrate geometry, enabling them to explore alternative network structures and the effect they have on the ANN behaviour. Although the work demonstrates the power of GAs to evolve networks of a particular topology, it still relies on choosing the type of network geometry beforehand and is therefore too constrained for our problem. The case for using novelty rather than an objective function as a guiding principle for exploring the feature space has been made by~\cite{Lehman2008} with the Novelty Search algorithm in which solutions evolved are less restricted. 

In this paper we explore the viability of using GAs to (a) generate networks that satisfy various requirements and (b) sample the diversity of the space of possible network structures under those requirements. In Section~\ref{sec::methods}, we describe the GA with particular focus on the multi-objective fitness function and novelty-biased population updating. In Section~\ref{sec::results}, we report results obtained when network sizes, degree distributions, clustering are varied and characterise the diversity of the feature space of valid networks. Finally, in Section~\ref{sec::conclusions}, we discuss a number of avenues for further development, in particular, in relation to network encoding and scaling up, and ways to improve novelty search through either adaptive mutation rate or different fitness functions. 

%========================================================================================================
\section{Methods}\label{sec::methods}
%========================================================================================================
Two objectives must be fulfilled: 
\begin{enumerate}
\item find valid networks, that is, networks that have the desired degree distribution and global clustering
\item optimise the diversity of these networks, where diversity is specified by a set of measures.
\end{enumerate}
Below, we describe the distinctive features of our approach which is otherwise very standard. 

\subsection{Encoding of the networks}\label{ssec::encoding}
Due to the unknown nature of the feature space, we chose to encode networks through their adjacency matrix $M$. An element $m_{i,j}$ of $M$ was set to $1$ (unweighted networks) if there was a connection between $i$ and $j$ and $i\neq j$ (no self-connection), $0$ otherwise. As we further limited ourselves to undirected networks, only the upper (or lower) triangle of the matrix needed to be stored yielding a genome size of $(N-1)(N-2)/2$ where $N$ was the number of nodes in the network. Encoding networks by their adjacency matrix makes the extraction of their network theoretic characteristics trivial. All measures described in the paper were implemented using the Brain Connectivity Toolbox~\cite{Rubinov:2010aa} in the Matlab environment. 

\subsection{Fitness}\label{ssec::fitness}
%Decide whether we want to say something about Pareto optimisation at this point
Multi-objective optimisation in GAs is a complex problem with many alternative methods, see~\cite{Konak:2006aa,Zitzler:1999aa} for example. Here, for simplicity, we used a weighted sum (or priori) approach~\cite{Konak:2006aa}. Namely, the total fitness F was a weighted sum of two fitness terms -- the constraint fitness CF (see Section~\ref{sssec::CF}) and the novelty fitness NF (see Section~\ref{sssec::NF}) -- with their respective weighting determined through experiments. The involvement of these fitness terms in the population update will be detailed in Section~\ref{ssec::population}. 

\subsubsection{Network constraining fitness (CF)}\label{sssec::CF}
Each evolved network was assessed in terms of its compliance with two constraints: a specified degree distribution (CF1) and a set global clustering coefficient (CF2). Once again, a weighted sum approach was used with each component evaluated as a distance. Both components were given equal weighting, except in the case of k-regular networks in which case it was found helpful to set a higher weight on CF1, reflecting the increased difficulty of finding k-regular networks. The CF value was then scaled in the range [0,1], 0 being the best fitness and corresponding to both CF1=0 and CF2=0. Thus CF acted as a gating or filtering mechanism. 

\subsubsection{Novelty encouraging fitness (NF)}\label{sssec::NF}
The novelty (or diversity) of an individual network was measured in terms of 7 structural measures: (1-2) Number of connections and density of the network. Although unlikely to change much due to our fixed network size in Normal and Poisson distributed-networks, changes in these measures could have significant effect on overall structure; (3) Average shortest path, a very common measurement that is important to network dynamics~\cite{Newman:2010aa}; (4-5) mean and range of local clustering. As valid networks are only specified by a global clustering coefficient, there can be significant difference in how clustering is distributed across the network~\cite{Ritchie:2014aa}; and (6-7) mean and range of nodal betweenness centrality, an indicator of how involved a node is in the shortest paths between all nodes in the network. 
The NF value of a network was computed from the difference between its measures and the average of those of valid networks previously evolved by the GA (see next Section), scaled in the range $[0,1]$. Once again, a weighted-sum approach was used. The NF value was subtracted from CF in order to reward diversity, i.e., at a given CF value, that network with the highest NF becomes the fittest (the smallest F -- which can be a negative value).   

% ================================================================================================================
% Table is placed here to make sure it appears on top of page 3. 
% This table is only used in Section 3 however. 
\begin{table*}
\centering
\caption{Number of valid networks for all configurations tested. Numbers in bold denote configurations for which more than one run was obtained. The actual number of runs (between 1 and 4) is provided in the text where appropriate. $<$50 denote configurations for which runtime exceeded a self-imposed limit of 2 days on the high-performance computing facilities. The exact number of valid networks is not shown as these configurations were excluded from any analysis. Empty cells denote configurations for which simulations are planned but yet to be carried out.  
}
\label{tab::configs}
\begin{tabular}{c|l|r|r|r|r|r|r|r|r|r|r|r|r|r|} 
\cline{2-15}
& Size & \multicolumn{4}{|c|}{12} & 16 & \multicolumn{4}{|c|}{20} & 60 & 110 & 160 & 200 \\
\cline{2-15}
& Clustering & 0.2 & 0.4 & 0.6 & 0.8 & 0.2 & 0.2 & 0.4 & 0.6 & 0.8 & 0.2 & 0.2 & 0.2 & 0.2 \\
\cline{1-15}
\multicolumn{1}{|c|}{\parbox[t]{2mm}{\multirow{3}{*}{\rotatebox[origin=c]{90}{Distrib.}}}} & Poisson & \textbf{50} & 50 & $<$50 & $<$50 & \textbf{50} & \textbf{50} & $<$50 & $<$50 & $<$50 & 50 & 50 & 50 & 50 \\
\multicolumn{1}{|c|}{} & Normal & 50 & 50 & \textbf{50} & $<$50 & $<$50 & 50 & 50 & 50 & $<$50 & 50 & 50 & 50 & 50 \\
\multicolumn{1}{|c|}{} & 5-Regular & \textbf{50} & \textbf{50} & \textbf{50} & 50 & $<$50 & $<$50 & $<$50 & 50 & $<$50 & & & & \\
%\multicolumn{1}{|c|}{} & 6-Regular & 50 & 50 & & & \textbf{50} & 50 & & & \textbf{50} & 50 & & & & & & \\
\cline{1-15}
\end{tabular}
\end{table*}
% ================================================================================================================

\subsection{Population updating}\label{ssec::population}
To mitigate the effects of increasing network size, we used a steady state GA~\cite{Noever:1992aa}, i.e., unlike the more traditional generational method, all individuals (100 in our implementation) remain in the population until they are replaced. Elitism in the population was introduced as follows. When valid networks (CF=0) appeared, they were removed (one individual per generation at most) from the population and placed in a separately maintained pool of solutions (of size 50 in our implementation). It was against those solutions that the NF fitness was calculated, i.e., new individuals were selected based on how different they were from previously evolved valid networks (NF=0 when the pool of solutions is empty). Although multiple copies of the same or similar networks were allowed into the pool of solutions, this only happened when these networks were extremely prevalent since they were actively discouraged by the GA through the novelty bias implemented by NF. Thus, the distribution of solutions in the pool could be thought of as a coarse approximation of the prevalence of structure types in the space of solutions (as long as prevalence is high). 

Note that since the pool of solutions changes, the process resembles a form of \textit{environmental evolution} with the fitness of the population affected by external factors that are then changed by the population -- albeit on a different time scale -- rather than a straightforward Pareto optimization involved in using more than one fitness measure. 

\subsection{Selection}
At the start of the GA the population was initialised by randomly assigning 1 or 0 to each element of the individual's genes. 
The (total) fitness F of each individual network in the population was calculated and then updated each time a mutation was applied to an individual.
As the fitness of an individual was only updated when it was selected, for the first few iterations after a network was added to the pool of solutions, those individuals in the better part of the population were still most likely to be selected despite being similar to the network just added to the pool. Although this is quickly corrected after a few iterations, it means that selection methods such as fitness-based selection (where a individual's chance of being selected is proportional to only its fitness) is not very effective at selecting potentially interesting individuals, i.e., individuals that are structurally different from the network just added to the pool of solutions, albeit with an as-yet uncompetitive fitness). Thus, we used rank-based selection in descending order (i.e., the best individual in the population is ranked last). 

\subsection{Genetic operations (mutation)}
Other than through replacing networks added to the pool of solution, mutation was the only source of creation of new networks. The mutation rate (normally 60\%) was set as the percentage of \textit{potential} links of a randomly picked node whose values (1 if the link exists, 0 otherwise) were flipped with probability 0.5. It should immediately be noted that such mutation does not preserve the degree distribution (there is no rewiring). Whilst it would be trivial to do so, this (computationally wasteful) approach was selected in first instance in order to maximise the exploration of the feature space.  Note that when dealing with k-regular networks, where all nodes must have the same number of connections, we found that even finding valid networks (CF=0) proved a significant challenge. In such cases, a slight variation was used which will be described in Section~\ref{sssec::Kregularnetworks}. 

%========================================================================================================
\section{Results}\label{sec::results}
%========================================================================================================

To establish the feasibility of (1) find valid networks and (2) maximise their diversity, we investigated the performance of our approach when sampling target configurations along two dimensions: network size and topological complexity. As shown in Table~\ref{tab::configs}, the network sizes considered varied between 12 and 200 nodes. Both homogeneous networks (k-regular networks) and heterogeneous (Poisson distributed, normally distributed) networks were evolved with the target global clustering ranging from 0.2 to 0.8. In the below, we report and analyse our main findings in terms of (i) the ability of our approach to produce valid networks with alternative structures (see Section~\ref{ssec::validnetworks}), (ii) the effectiveness of the approach in controlling the diversity of the population of valid networks obtained (see Section~\ref{ssec::diversity}), and (iii) the possibility that the above results provide useful insights as to the structure of the space of valid networks. 

\begin{figure}
\centering
\includegraphics[width=0.32\textwidth]{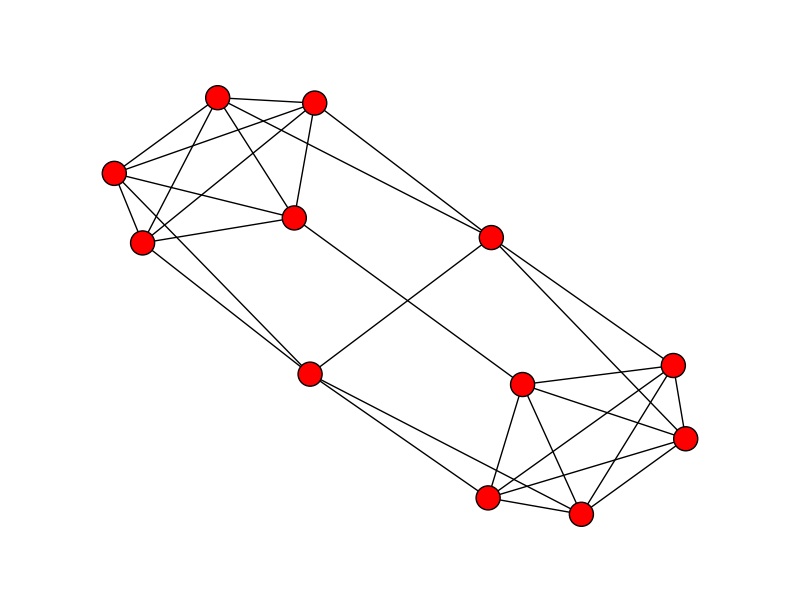}
\includegraphics[width=0.32\textwidth]{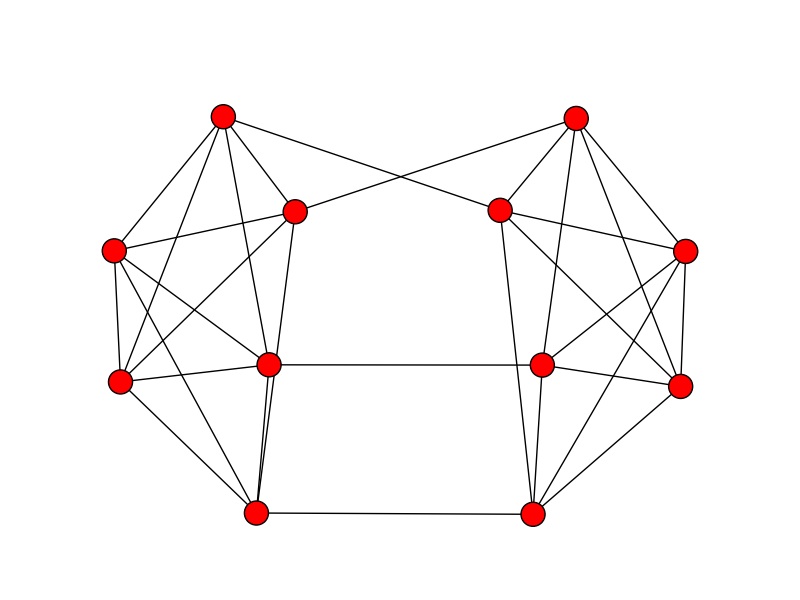}
\includegraphics[width=0.32\textwidth]{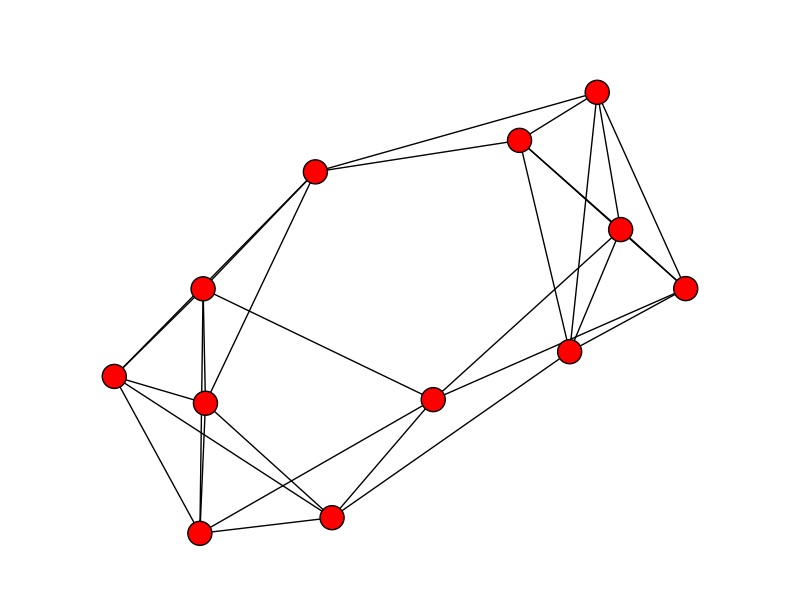}
\caption{Three networks from the pool of solutions obtained from one run for 5-regular networks with 12 nodes and global clustering C=0.6.} 
\label{fig::fig1kregular}
\end{figure}

\begin{figure*}
\centering
\includegraphics[width=0.32\textwidth]{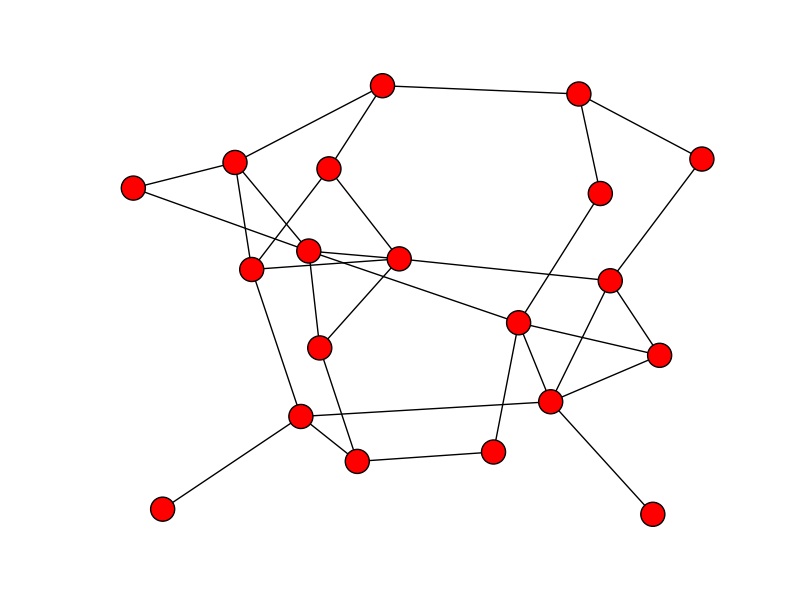}
\includegraphics[width=0.32\textwidth]{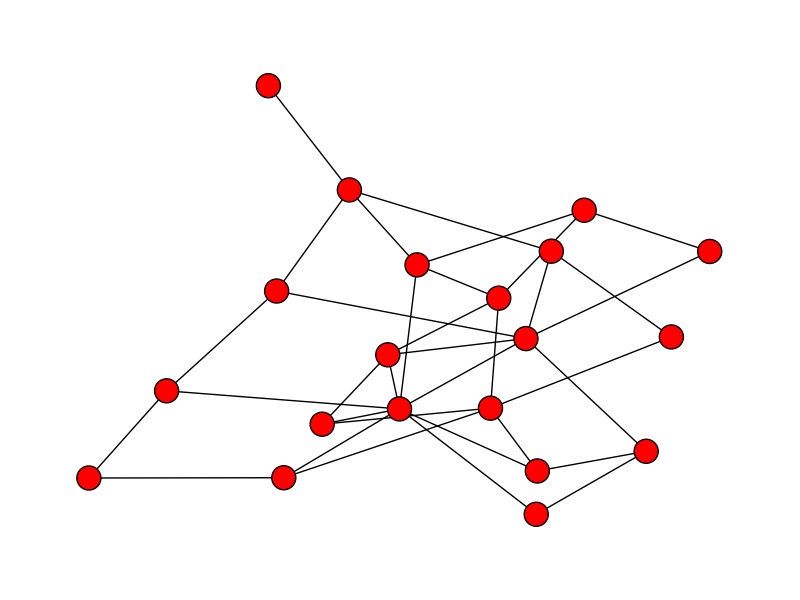}
\includegraphics[width=0.32\textwidth]{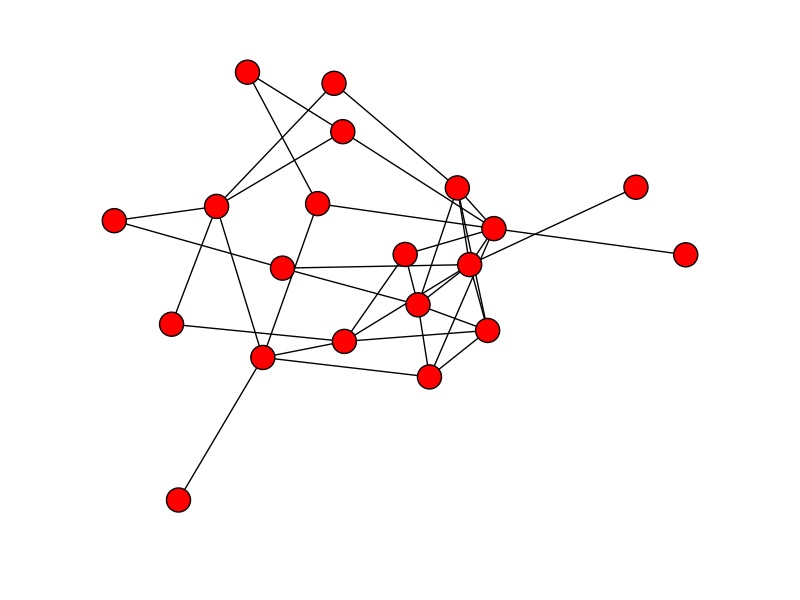}
\includegraphics[width=0.32\textwidth]{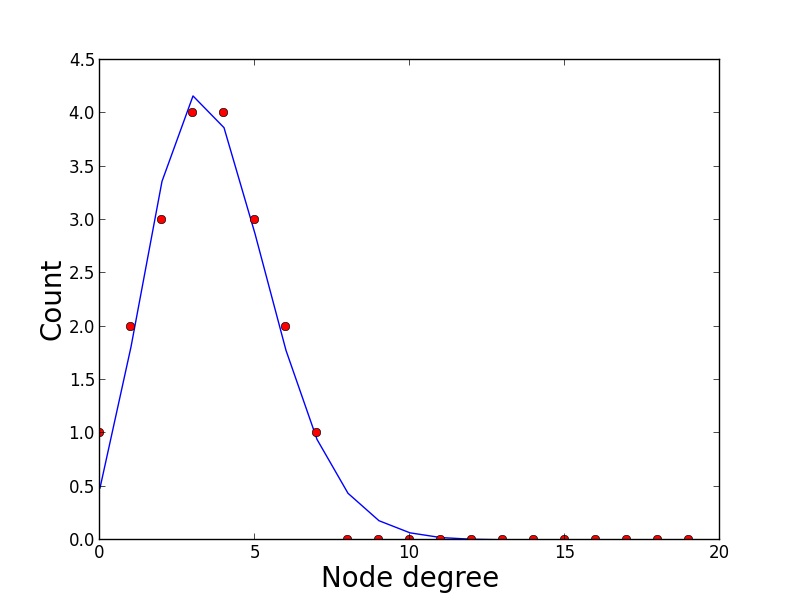}
\includegraphics[width=0.32\textwidth]{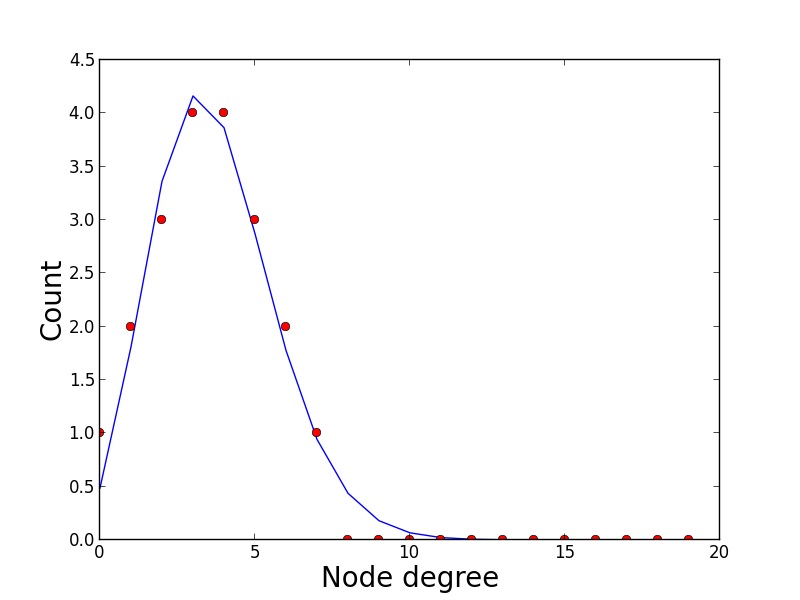}
\includegraphics[width=0.32\textwidth]{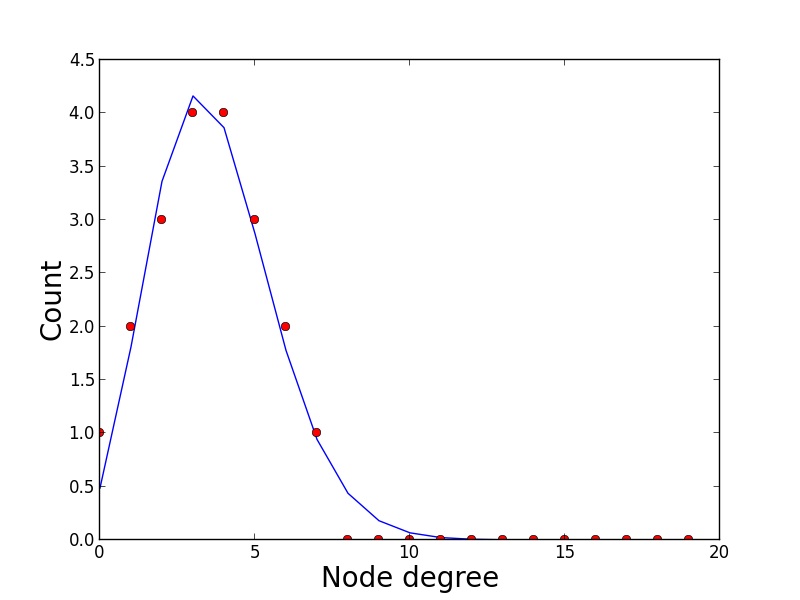}
\caption{Three networks from the pool of solutions obtained from one run for Poisson networks with 20 nodes and global clustering C=0.2.} 
\label{fig::fig1poisson}
\end{figure*}

%==============================================================================
\subsection{Ability to produce valid networks with diverse structure}\label{ssec::validnetworks}
\subsubsection{K-regular networks}\label{sssec::Kregularnetworks}
We begin with small networks with a homogeneous degree distribution. K-regular networks are networks in which all nodes have the same degree (number of links).  For example, a 2-regular network would be a loop where all nodes have exactly 2 connections~\cite{Meringer:1999aa}. This enables us to tightly control the overall density of the network (value of K divided by the number of nodes N) whilst implementing a degree distribution which is very simple to test for. 
The CF1 fitness of the k-regular networks was determined as the sum of two terms: (1) a regularity term calculated as the range of degrees found in the network -- with 0 denoting a regular network, and (2) a degree term calculated as the difference between target and measured degrees. Of these 2 terms, the most challenging for the GA to minimise is the regularity. This is because whereas a single flip (creation or deletion of a link) will uniquely determine whether the degree term increases or decreases, there is no mechanism to prevent this flip from negatively impacting the regularity term (e.g., through deleting a link from the node with the smallest degree). To (partly) mitigate this problem without biasing the novelty of the search, the CF1 component of the CF fitness was weighted higher than CF2 (CF1=60\% of total CF) to favour establishment of the regularity of the network before searching for the desired global clustering level. 

Here, it is worth mentioning that when small-sized k-regular networks are considered, not all configurations can be realised, e.g., a 5-regular network with 13 nodes. This is further compounded by the constraint of a set global clustering coefficient. We therefore limited our investigations to possible configurations, as given in~\cite{Meringer:1999aa}, with preliminary experiments used to determine realisable clustering coefficients by testing whether it was possible to find at least one valid network. We found that when limiting CF1 to its regularity component, it was always possible to find networks achieving a set level of global clustering. 

Figure~\ref{fig::fig1kregular} shows three networks from the pool of solutions for a representative configuration. It demonstrates that (a) it was possible to evolve valid networks and (b) that there was some diversity in the pool of solutions. Examination of the time course for the evolution of one valid k-regular network (data not shown) revealed that as the value of CF1 decreased the time needed to achieve this decrement increased slightly. This trend continued until the regularity component of CF1 reached 1, i.e., within $\pm 1$ from the desired value of K, at which point the time taken increased sharply. This is easily explained by the much reduced likelihood of a beneficial mutation. To address this, we modified the mutation mechanism so that when an individual with CF1 $\leq$ 2 was subjected to mutation the node selected to be mutated was chosen via rank-based selection, with the nodes with the most common degree value least likely to be selected. This led to a reduction in the time taken without any effect on the observed level of diversity (data not shown). This method was only applied to k-regular networks and would not be applicable to any other form of degree distribution without significantly affecting diversity. 

\subsubsection{Poisson and normal networks}\label{sssec::othernetworks}
As k-regular networks are hardly representative of real-world networks, we now consider two other forms of degree distribution: Poisson and normal. Note that whereas Poisson distributed networks are frequently studied -- they are quite common in telecommunication, astronomy, biology and large random networks~\cite{Newman:2010aa}, networks with normal degree distribution are not. This distribution was purely selected for the purpose of demonstrating the generality of our approach. A more realistic distribution would be either scale free or log-normal. 

For both types of network, the CF1 value was calculated on the basis of the $R^2$ value obtained when fitting the network's degree distribution with the expected distribution. CF1 was set to 0 when R$^2$<0.05. We should acknowledge that an oversight led to the average degree not being actually controlled for. However, we found that the evolved networks tended to have a remarkably similar average degree, and, where necessary, relevant quantities (e.g., number of motifs) were normalised (see Section~\ref{ssec::higherorderstructure} for example) so that we do not believe this oversight affects the validity of our results in any way. 

Figures~\ref{fig::fig1poisson} and~\ref{fig::fig1normal} show three networks from the pool of solutions for representative configurations of Poisson and normal networks. These figures confirm that yet again, (a) it was possible to evolve valid networks and (b) that there was diversity in the pool of solutions. Runtime for these types of network was much shorter than for k-regular networks and network sizes up to 200 nodes were attempted with similar results (data not shown due to difficulty with showing structure in large networks). 

\begin{figure*}
\centering
\includegraphics[width=0.32\textwidth]{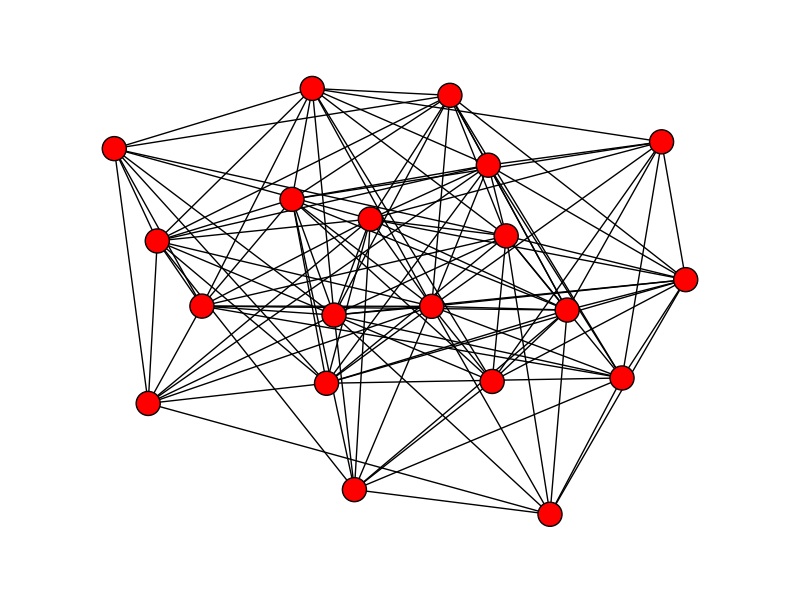}
\includegraphics[width=0.32\textwidth]{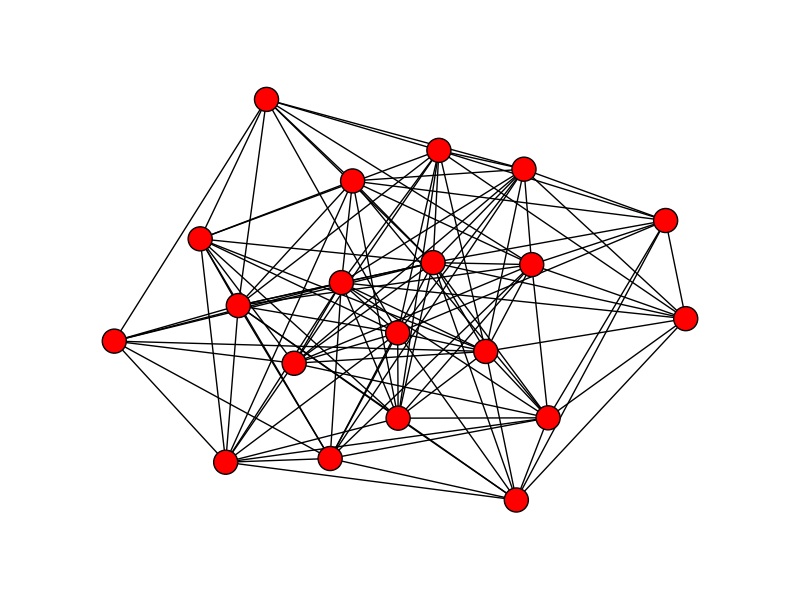}
\includegraphics[width=0.32\textwidth]{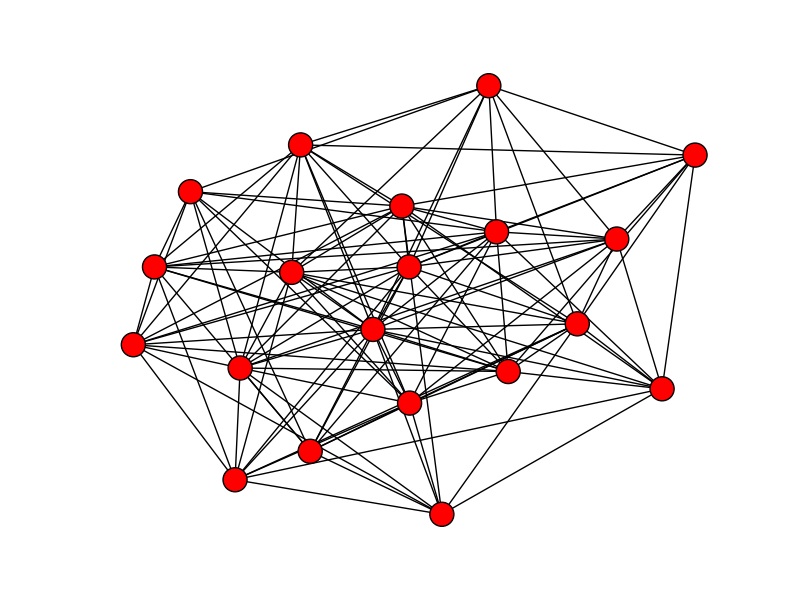}
\includegraphics[width=0.32\textwidth]{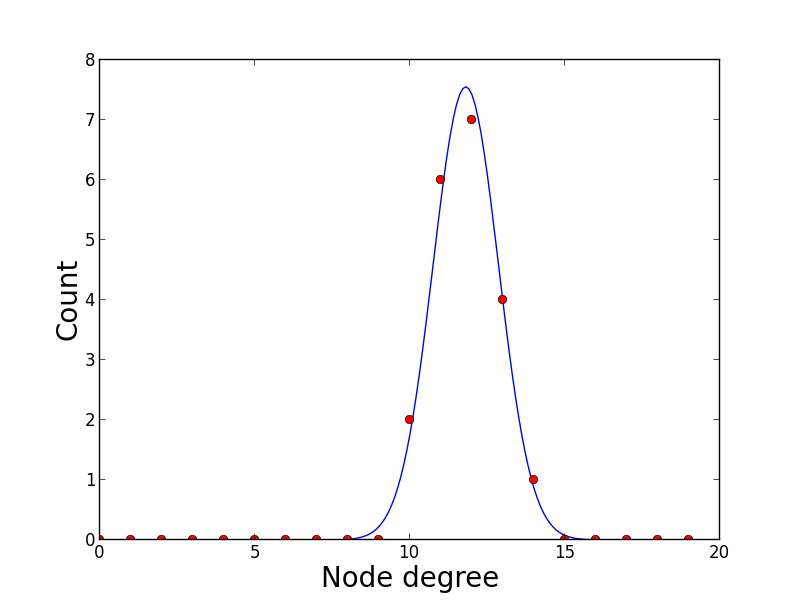}
\includegraphics[width=0.32\textwidth]{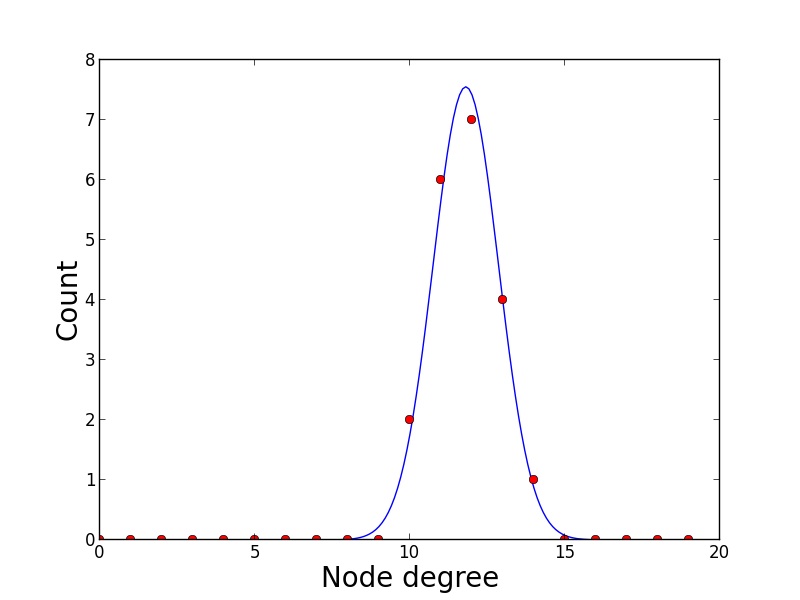}
\includegraphics[width=0.32\textwidth]{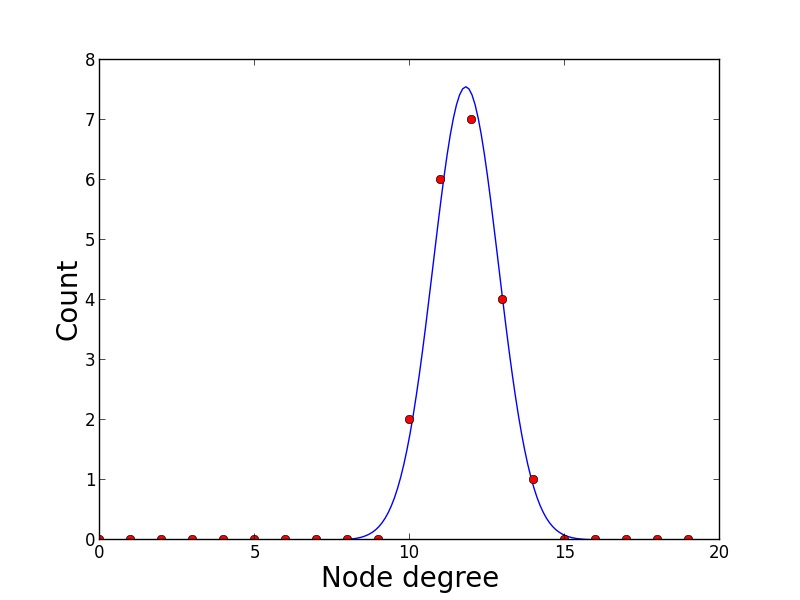}
\caption{Three networks from the pool of solutions obtained from one run for normal networks with 20 nodes and global clustering C=0.6.} 
\label{fig::fig1normal}
\end{figure*}

%\newpage
\subsubsection{A note about runtime}
We limited all runs to two days of runtime on the high-performance computing cluster (64 cores and 256GB RAM). As a result, we could not obtain the required 50 valid networks in a number of configurations (see Table~\ref{tab::configs}). As previously mentioned, because we were mostly concerned with establishing proof of principle, we adopted simple but clearly wasteful forms of network encoding and mutation. Although it is clear that longer runtimes were to be expected from increasing network size (genome size increasing in N$^2$), the complexity of the space of possible solutions also plays a critical role and cannot be underestimated. For example, at equal network size (12-20 nodes), there were significant differences in runtime between k-regular networks on the one hand and Poisson and normal networks on the other hand, the former being much costlier. This aspect will be investigated further in Section~\ref{ssec::landscape}. 

%==============================================================================
\subsection{Effectiveness of the novelty bias}\label{ssec::diversity}
Here we consider the effectiveness of our approach in controlling diversity in the pool of valid networks. We start by noting that, except for the most trivial of all setups (e.g., very small k-regular networks for which the entire space of solution is computationally tractable), it is impossible to quantify the extent to which the 50 evolved networks are representative of the true space of possible realisations for each of the configuration considered. Therefore, in the absence of ground truth, we examine the effect that selection for novelty has on the diversity of the networks evolved. We carried out simulations for 3 conditions. In the first condition (NF:off), the NF value was held to 0 throughout evolution, i.e., no novelty bias. In the second condition (NF:on), the NF value was calculated and used as described in Sections~\ref{ssec::fitness} and~\ref{ssec::population}. Finally, in the third condition (NF:Loc Clus), only the range of local clustering (item 5 in the definition of NF, see Section~\ref{ssec::fitness}) was used to determine NF. As shown by Figure~\ref{fig::noveltyconditions}, without the novelty bias, the pool of networks show less diversity (over the 7 measures) and across all distributions. This decrease in diversity is not uniform, with normally distributed networks showing the largest effect. 

\begin{figure*}
\centering
\includegraphics[width=.9\linewidth]{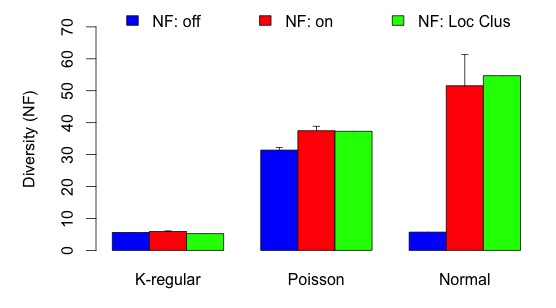}
\caption{Diversity in k-regular (K=5, C=0.6; 4 runs for the NF:on condition,  1 run otherwise), Poisson (C=0.2; 3 runs for the NF:on condition, 2 runs for the NF:off condition, 1 otherwise) and normal (C=0.6; 2 runs for the NF:on condition, 1 otherwise) 12-node networks when three novelty scenarios are considered: NF:off (blue), no novelty fitness involved in the population update; NF:on (red), the default mode of operation; NF:Loc Clus (green), NF calculated as the range of local clustering. Standard deviations are shown where available.}
\label{fig::noveltyconditions}
\end{figure*}

These differences suggest that although some diversity is possible without the selective pressure implemented by NF, it is limited. When NF is determined by the range of local clustering, diversity is comparable (or even greater in the case of normally-distributed networks) to that when all 7 items are included. This suggests that of the 7 measures, maximising the range of local clustering may be one of the most effective means of promoting diversity in the population. This is not particularly surprising as the range of local clustering is a reasonable marker of the diversity of motif distribution at node level (see Section~\ref{ssec::higherorderstructure} for more on this topic). However, this is speculative only and it could also be the result of the GA reaching a Pareto optimum when trying to fulfil all NF measures. More thorough testing is required to determine how sensitive diversity is to each of the 7 measures. 

%=======================================================================
\subsection{Insights about the space of solutions}\label{ssec::landscape}
An interesting notion is that performance of the GA in evolving a population of valid solutions may provide useful insights as to the nature of the space of solutions. In this Section, we examine the diversity of the evolved networks in terms of how it is affected by changes in degree distribution, global clustering, network size and mutation rate, as well as in terms of complexity beyond that specified by the fitness function.

\begin{figure*}
\centering
\includegraphics[width=.9\linewidth]{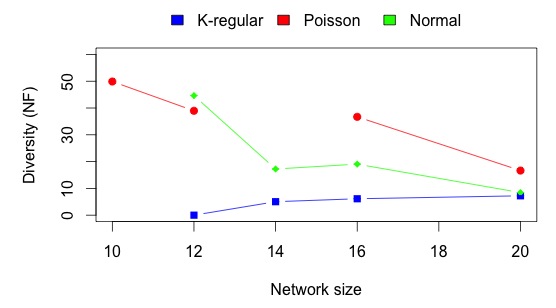}
\caption{Diversity in k-regular (K=6, C=0.2; 1 run), Poisson (C=0.2; 3 runs for N=12, 1 otherwise) and normal (C=0.6; 2 runs for N=12, 1 otherwise) networks when network size is varied. Standard deviations are shown where available although too small to be visible.}
\label{fig::fig9}
\end{figure*}

%==============================================================================
\subsubsection{Degree distribution, clustering and diversity}
We begin by considering the effect on diversity of increasing the network size. Here, the intuition is that an increased network size should lead to an increase in the number of valid networks possible, potentially leading to an increase in diversity of the valid networks. Figure~\ref{fig::fig9} shows the diversity obtained when varying the network size from 10 to 20 (data for N$>$20 not shown) for the three distributions considered. For k-regular networks, there is a small but significant increase in diversity with increasing node size. Although the small range of network sizes considered to date means we can only speculate as to whether this trend would continue, the low diversity (in relation to the other two distributions) is consistent with the fact that in a strictly controlled degree distribution such as with k-regular networks, alternative networks may be few. In fact it is highly likely that with the pool size used here, the majority of network structures possible have been included. Support for this hypothesis comes from Figure~\ref{fig::noveltyconditions} in which it is observed that the diversity of the pool of networks is barely affected by the inclusion of the novelty bias in the selective pressure. Further evidence comes from Figure~\ref{fig::fig13} and will be discussed in Section~\ref{ssec::sparsity}. Finally, the lack of alternative network structures is demonstrated by how diversity was affected by a change in the target global clustering (data not shown). Diversity was unchanged when clustering values were 0.2, 0.4 and 0.6. When clustering reached 0.8 (which is very prescriptive), diversity collapsed to near 0 value. When non-trivial clustering coefficients are considered, the number of possible networks increase with larger network sizes, however, diversity in the pool of solutions only marginally increases. 

\begin{figure*}
\centering
\includegraphics[width=.9\linewidth]{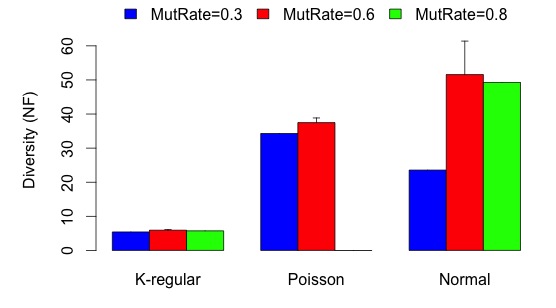}
\caption{Diversity (as measured by the NF fitness) for k-regular (K=5, C=0.2; 4 runs for mutation rate of 0.6, 1 run otherwise), Poisson (C=0.2; 2 runs for mutation rate of 0.6, 1 run otherwise) and normal (C=0.6; 2 runs for mutation rate of 0.6, 1 run otherwise) networks with 12 nodes. For each network configuration, data is provided for mutation rate of 0.3, 0.6 and 0.8 (from left to right). Standard deviations are shown where available.}
\label{fig::fig13}
\end{figure*}

With Poisson and normal networks, a complete reversal of the trend is observed with a decrease in diversity with increasing network size. This decrease continues when larger network sizes are considered (from 60 to 200 nodes, data not shown). This decrease is associated with a range of local clustering which converges (data not shown). Although this finding may seem counter-intuitive, this result is consistent with previous findings suggesting that as networks become larger the complexity of their network diminishes. For example, loops, which play a key role in the emergence of higher-order structure (see Section~\ref{ssec::higherorderstructure}) are only seen in relatively small networks~\cite{Boccaletti:2006aa}. When the target global clustering coefficient increases, diversity decreases (data not shown). Once again, it is observed that the range of local clustering converges to a fixed value. This is expected since increased clustering imposes tighter constraints on the types of sub-graph a node can be associated with. Namely, the number of closed triples should increase which can only be implemented through closed order-3 subgraphs (i.e., triangles) or order-4 subgraphs such as fully connected squares and squares with one diagonal (see Section~\ref{ssec::higherorderstructure}). Not surprisingly, it is the mean path length (item 3 in the definition of NF, see Section~\ref{ssec::fitness}) which shows the most variability. 

For all configurations considered, diversity in the pool of solutions for both Poisson and normal configurations exceeds that of k-regular networks, 
with normal networks showing the greatest diversity. Whether the difference uncovered by the GA is a true reflection of the space of possible solutions or merely a consequence of an increased likelihood of the GA actually finding solutions (e.g., due to the random graph generation used to initialise the networks) will be the subject of further investigation. 

%==============================================================================
\subsubsection{Sparsity of the space of solutions}\label{ssec::sparsity}
As the pool of valid networks is of fixed size, and it allows the addition of networks that are only slightly different to those already added, we can expect the rate of mutation to significantly impact how the GA explores the space of solutions. Very low rates should provide good sampling but at the cost of limited diversity. Very high rates, on the other hand, would lead to a random search and would likely be very low-yield, especially given our wasteful approach to mutation (here, rewiring would have a significant advantage). As previously stated, most of our results were obtained with a mutation rate of 0.6 as preliminary experiments showed it to be effective in most cases. Here, we examine the effect of a changing mutation rate on the diversity of the pool of valid networks. 

As shown by Figure~\ref{fig::fig13}, this effect is heavily dependent on the network's target distribution. For 5-regular 12-node networks, the rate of mutation has very little effect, consistent with our previous observations that there are only a few possible valid networks, clustered together in the feature space. In contrast, for normally-distributed networks, an increase of the mutation rate from 0.3 to 0.6 yields a considerable improvement in diversity suggesting that clusters of valid networks are scattered across the feature space and can only be accessed if the mutation rate is sufficiently high. When the mutation rate increases further (approaching random search), diversity starts decreasing.

%==============================================================================
\subsubsection{Beyond NF: Higher-order structure}\label{ssec::higherorderstructure}
Network analysis in a number of biological systems has revealed the presence of motifs, that is, subgraphs that occur at a frequency significantly higher than that expected by chance. Increased diversity in terms of motif distribution at node-level (i.e., the probabilities of a node to be a member of a set of motifs) should lead to increased range of local clustering (which we suggested played a key role in the diversity of the networks presented earlier). Here, we focus on normally-distributed networks for which measured diversity was maximal when NF was calculated on the basis of the range of local clustering (see Figure~\ref{fig::noveltyconditions}). For each network of a representative sample of 11 networks from the pool of valid 12-node networks with global clustering of 0.6, we determined the prevalence of 5 order-4 motifs (fully connected squares, squares with one diagonal, empty squares, stars and open quadruples) using the motif-counting algorithm proposed in~\cite{Ritchie:2014aa}. To account for average degree variation between networks, the count for each motif was first normalised to the number of triangles in the network (since all networks are valid, the ratio of triangles to open triples was fixed and equal to 0.6). As shown by Figure~\ref{fig::fig14}, substantial diversity in motif composition is observed despite the preserved global clustering.
 
\begin{figure}
\centering
\includegraphics[width=.9\linewidth]{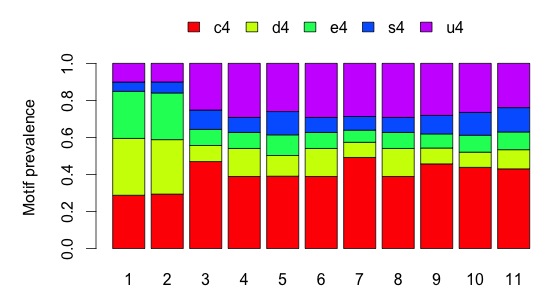}
\caption{Prevalence of order-4 motifs in a sample of valid networks found for 12-node networks with normal degree distribution and a global coefficient of 0.6. Motifs are as follows: fully connected square (c4), square with one diagonal (d4), square (e4), star (s4), open chain (u4).}
\label{fig::fig14}
\end{figure} 

%========================================================================================================
\section{Conclusions}\label{sec::conclusions}
%========================================================================================================

Being capable of generating networks with given network theoretic characteristics would be very useful at a time when there is strong focus on a network-centric analysis of complex systems. Indeed, an essential tool for inference is the availability of proper null models. However, the current state of the art is quite limited, with available mathematical methods only capable of dealing with a few coarse indicators (typically degree distribution and global clustering). Even then, there is currently no mechanism by which to explore the diversity of the space of solutions for a given set of constraints. With this paper, we set on exploring the possibility that GAs provide a viable alternative. We described a novelty-biased methodology which we tested in a range of scenarios (different degree distributions and global clustering coefficients). Our results suggest that GAs are indeed a potentially viable approach to sampling the space of solutions of networks satisfying network theoretic characteristics. As far as we know, our paper is the first proof of concept in this area. 

However, although we have shown that the GA was capable of finding novel valid networks in a range of degree distributions and global clustering, and for a range of sizes, there were cases (e.g., Poisson distributed networks with clustering of 0.8 or higher) where the GA was unable to find 50 valid networks within a reasonable time frame, namely, two days of runtime on the high-performance computing cluster (64 cores and 256GB RAM). 
Whilst this can be possibly attributed to the sparse nature of the space of valid networks, it is also clear that our methodology could be improved at various levels. In the following, we identify three key avenues for further development. 

\subsubsection*{Encoding for larger networks}
It is quite manifest that, whilst an adjacency-matrix based encoding provides maximal accuracy, it is an inefficient representation that makes scaling to larger network sizes difficult, and our empirical resuts show that. From a theoretical viewpoint, however, it is unclear that larger networks should necessarily lead to more complex spaces of solutions. For example, in random networks, loops (which contribute to higher-order structure) disappear in the limit of network size. This leaves us with the interesting option of \textit{decreasing} the level of detail the GA uses for encoding the networks. Based on recent work~\cite{Ritchie:2014ab}, a possible solution would be to encode networks in terms of a motif decomposition. Mutations would then change the number and position of those motifs. In addition to decreasing the size of the genome, this method may allow the GA to focus on those motifs that most affect the diversity of the network. 

\subsubsection*{Adaptive mutation rate}
Our results have shown that the choice of rate of mutation greatly impacts both the speed of evolution and the diversity of solutions obtained. Given a fairly limited pool size and the lack of prior knowledge about the number of solutions to a given combination of constraints, it is of importance to try to ensure maximum diversity within the pool. In this paper, the mutation rate was set so that a connection of a selected node had a 60$\%$ chance of changing. 
This was found to be a good fit for all network sizes used here but it should be improved with further testing and/or knowledge of the feature space. For example, if valid networks are sparse but clustered, then it would be desirable to have a rate of mutation that is based on how far the individual being mutated is from one of these groups of valid networks. Control of the mutation rate would make it possible to control the tendency of new individuals to either explore the neighbourhood of a group of valid networks or maximize novelty.

\subsubsection*{Weighting of the fitness functions and alternative multi-objective optimization}
We chose weighted sum (or priori) approach to multi-objective optimisation for simplicity, and also because we had no prior knowledge about the structure of the space of solutions. However, this approach suffers from an important limitation in that the choice of weighting factors can significantly affect the results~\cite{Konak:2006aa}. In this paper, we made do with a fixed weight for each of factor, estimated from preliminary experiments. This, along with the changing rate of NF as the GA unfolded was sufficient to coarsely sample the space of possible valid networks and guess their diversity. Use of methods more similar to random-weight or weight-based GA~\cite{Hajela:1992aa} may enable us to say more about the detailed topology of the space of valid networks. Alternatively, one could use a more Pareto set like method whereby a population of valid networks is evolved and then those networks are explored for novelty via an altering objective function such as VEGA~\cite{Schaffer:1985aa} or some kind of Pareto-reckoning approach such as SPEA~\cite{Zitzler:1999aa}. In particular, if valid networks are clustered, one could evolve valid networks via a method of "niched-selection"~\cite{Jimenez:2002aa}.

{\ifthenelse{\boolean{publ}}{\footnotesize}{\small}
\bibliographystyle{bmc_article}
\bibliography{bibliography}  % sigproc.bib is the name of the Bibliography in this case


\begin{thebibliography}{10}
\providecommand{\url}[1]{[#1]}
\providecommand{\urlprefix}{}

\bibitem{Sporns2005}
Sporns O, Tononi G, K\"{o}tter R: \textbf{{The human connectome: A structural
  description of the human brain.}} \emph{PLoS Computational Biology} 2005,
  \textbf{1}(4):e42.

\bibitem{Bansal:2009aa}
Bansal S, Khandelwal S, Meyers L: \textbf{Exploring biological network
  structure with clustered random networks}. \emph{BMC Bioinformatics} 2009,
  \textbf{10}:405.

\bibitem{Newman:2010aa}
Newman M: \emph{Networks: An introduction}. Oxford University Press 2010.

\bibitem{House2010}
House T, Keeling MJ: \textbf{{The impact of contact tracing in clustered
  populations.}} \emph{PLoS computational biology} 2010,
  \textbf{6}(3):e1000721.

\bibitem{Green2010}
Green DM, Kiss IZ: \textbf{{Large-scale properties of clustered networks:
  implications for disease dynamics.}} \emph{Journal of biological dynamics}
  2010, \textbf{4}(5):431--45.

\bibitem{Ritchie:2014ab}
{Ritchie} M, {Berthouze} L, {Kiss} IZ: \textbf{{Beyond clustering: Mean-field
  dynamics on networks with arbitrary subgraph composition}}. \emph{ArXiv
  e-prints} 2014.

\bibitem{DelGenio2010}
{Del Genio} CI, Kim H, Toroczkai Z, Bassler KE: \textbf{{Efficient and exact
  sampling of simple graphs with given arbitrary degree sequence.}} \emph{PloS
  one} 2010, \textbf{5}(4):e10012.

\bibitem{Milo2003}
Milo R, Kashtan N, Itzkovitz S, Newman MEJ, Alon U: \textbf{{Uniform generation
  of random graphs with arbitrary degree sequences}}. \emph{Arxiv preprint
  condmat0312028} 2003, \textbf{cond-mat/0}:1--4,
  \urlprefix\url{[http://arxiv.org/abs/cond-mat/0312028]}.

\bibitem{Ritchie:2014aa}
Ritchie M, Berthouze L, House T, Kiss IZ: \textbf{Higher-order structure and
  epidemic dynamics in clustered networks}. \emph{Journal of Theoretical
  Biology} 2014, \textbf{348}:21--32.

\bibitem{Konak:2006aa}
Konak A, Coit DW, Smith AE: \textbf{Multi-objective optimization using genetic
  algorithms: A tutorial}. \emph{Reliability Engineering \& System Safety}
  2006, \textbf{91}(9):992--1007.

\bibitem{Risi:2012aa}
Risi S, Stanley KO: \textbf{An enhanced hypercube-based encoding for evolving
  the placement, density, and connectivity of neurons}. \emph{Artificial life}
  2012, \textbf{18}(4):331--363.

\bibitem{Lehman2008}
Lehman J, Stanley KO: \textbf{{Exploiting Open-Endedness to Solve Problems
  Through the Search for Novelty}}. In \emph{Artificial Life XI} 2008:329--336.

\bibitem{Rubinov:2010aa}
Rubinov M, Sporns O: \textbf{Complex network measures of brain connectivity:
  Uses and interpretations}. \emph{Neuroimage} 2010, \textbf{52}(3):1059--1069.

\bibitem{Zitzler:1999aa}
Zitzler E, Thiele L: \textbf{Multiobjective evolutionary algorithms: a
  comparative case study and the strength Pareto approach}. \emph{IEEE
  Transactions on Evolutionary Computation} 1999, \textbf{3}(4):257--271.

\bibitem{Noever:1992aa}
Noever D, Baskaran S: \textbf{Steady state vs. generational genetic algorithms:
  A comparison of time complexity and convergence properties}. SFI Working
  Paper 1992-07-032, Santa Fe Institute 1992.

\bibitem{Meringer:1999aa}
Meringer M: \textbf{Fast generation of regular graphs and construction of
  cages}. \emph{Journal of Graph Theory} 1999, \textbf{30}(2):137--146.

\bibitem{Boccaletti:2006aa}
Boccaletti S, Latora V, Moreno Y, Chavez M, Hwang DU: \textbf{Complex networks:
  Structure and dynamics}. \emph{Physics reports} 2006,
  \textbf{424}(4):175--308.

\bibitem{Hajela:1992aa}
Hajela P, Lin CY: \textbf{Genetic search strategies in multicriterion optimal
  design}. \emph{Structural Optimization} 1992, \textbf{4}(2):99--107.

\bibitem{Schaffer:1985aa}
Schaffer JD: \textbf{Multiple objective optimization with vector evaluated
  genetic algorithms}. In \emph{Proceedings of the International Conference on
  Genetic Algorithm and their Applications}. Edited by Grefenstette JJ
  1985:93--100.

\bibitem{Jimenez:2002aa}
Jimenez F, Gomez-Skarmeta A, Sanchez G, Deb K: \textbf{An evolutionary
  algorithm for constrained multi-objective optimization}. In \emph{Proceedings
  of the 2002 Congress on Evolutionary Computation}, \emph{Volume~2}, IEEE
  2002:1133--1138.

\end{thebibliography}

\newcommand{\BMCxmlcomment}[1]{}

\BMCxmlcomment{

<refgrp>

<bibl id="B1">
  <title><p>{The human connectome: A structural description of the human
  brain.}</p></title>
  <aug>
    <au><snm>Sporns</snm><fnm>O</fnm></au>
    <au><snm>Tononi</snm><fnm>G</fnm></au>
    <au><snm>K\"{o}tter</snm><fnm>R</fnm></au>
  </aug>
  <source>PLoS Computational Biology</source>
  <pubdate>2005</pubdate>
  <volume>1</volume>
  <issue>4</issue>
  <fpage>e42</fpage>
</bibl>

<bibl id="B2">
  <title><p>Exploring biological network structure with clustered random
  networks</p></title>
  <aug>
    <au><snm>Bansal</snm><fnm>S.</fnm></au>
    <au><snm>Khandelwal</snm><fnm>S.</fnm></au>
    <au><snm>Meyers</snm><fnm>L.</fnm></au>
  </aug>
  <source>BMC Bioinformatics</source>
  <pubdate>2009</pubdate>
  <volume>10</volume>
  <issue>1</issue>
  <fpage>405</fpage>
</bibl>

<bibl id="B3">
  <title><p>Networks: An introduction</p></title>
  <aug>
    <au><snm>Newman</snm><fnm>M</fnm></au>
  </aug>
  <publisher>Oxford University Press</publisher>
  <pubdate>2010</pubdate>
</bibl>

<bibl id="B4">
  <title><p>{The impact of contact tracing in clustered
  populations.}</p></title>
  <aug>
    <au><snm>House</snm><fnm>T</fnm></au>
    <au><snm>Keeling</snm><fnm>MJ</fnm></au>
  </aug>
  <source>PLoS computational biology</source>
  <pubdate>2010</pubdate>
  <volume>6</volume>
  <issue>3</issue>
  <fpage>e1000721</fpage>
</bibl>

<bibl id="B5">
  <title><p>{Large-scale properties of clustered networks: implications for
  disease dynamics.}</p></title>
  <aug>
    <au><snm>Green</snm><fnm>DM</fnm></au>
    <au><snm>Kiss</snm><fnm>IZ</fnm></au>
  </aug>
  <source>Journal of biological dynamics</source>
  <pubdate>2010</pubdate>
  <volume>4</volume>
  <issue>5</issue>
  <fpage>431</fpage>
  <lpage>-45</lpage>
</bibl>

<bibl id="B6">
  <title><p>{Beyond clustering: Mean-field dynamics on networks with arbitrary
  subgraph composition}</p></title>
  <aug>
    <au><snm>{Ritchie}</snm><fnm>M.</fnm></au>
    <au><snm>{Berthouze}</snm><fnm>L.</fnm></au>
    <au><snm>{Kiss}</snm><fnm>I. Z.</fnm></au>
  </aug>
  <source>ArXiv e-prints</source>
  <pubdate>2014</pubdate>
</bibl>

<bibl id="B7">
  <title><p>{Efficient and exact sampling of simple graphs with given arbitrary
  degree sequence.}</p></title>
  <aug>
    <au><snm>{Del Genio}</snm><fnm>CI</fnm></au>
    <au><snm>Kim</snm><fnm>H</fnm></au>
    <au><snm>Toroczkai</snm><fnm>Z</fnm></au>
    <au><snm>Bassler</snm><fnm>KE</fnm></au>
  </aug>
  <source>PloS one</source>
  <pubdate>2010</pubdate>
  <volume>5</volume>
  <issue>4</issue>
  <fpage>e10012</fpage>
</bibl>

<bibl id="B8">
  <title><p>{Uniform generation of random graphs with arbitrary degree
  sequences}</p></title>
  <aug>
    <au><snm>Milo</snm><fnm>R</fnm></au>
    <au><snm>Kashtan</snm><fnm>N</fnm></au>
    <au><snm>Itzkovitz</snm><fnm>S</fnm></au>
    <au><snm>Newman</snm><fnm>M E J</fnm></au>
    <au><snm>Alon</snm><fnm>U</fnm></au>
  </aug>
  <source>Arxiv preprint condmat0312028</source>
  <pubdate>2003</pubdate>
  <volume>cond-mat/0</volume>
  <fpage>1</fpage>
  <lpage>-4</lpage>
  <url>http://arxiv.org/abs/cond-mat/0312028</url>
</bibl>

<bibl id="B9">
  <title><p>Higher-order structure and epidemic dynamics in clustered
  networks</p></title>
  <aug>
    <au><snm>Ritchie</snm><fnm>M</fnm></au>
    <au><snm>Berthouze</snm><fnm>L</fnm></au>
    <au><snm>House</snm><fnm>T</fnm></au>
    <au><snm>Kiss</snm><fnm>IZ</fnm></au>
  </aug>
  <source>Journal of Theoretical Biology</source>
  <pubdate>2014</pubdate>
  <volume>348</volume>
  <fpage>21</fpage>
  <lpage>32</lpage>
</bibl>

<bibl id="B10">
  <title><p>Multi-objective optimization using genetic algorithms: A
  tutorial</p></title>
  <aug>
    <au><snm>Konak</snm><fnm>A</fnm></au>
    <au><snm>Coit</snm><fnm>DW</fnm></au>
    <au><snm>Smith</snm><fnm>AE</fnm></au>
  </aug>
  <source>Reliability Engineering \& System Safety</source>
  <pubdate>2006</pubdate>
  <volume>91</volume>
  <issue>9</issue>
  <fpage>992</fpage>
  <lpage>1007</lpage>
</bibl>

<bibl id="B11">
  <title><p>An enhanced hypercube-based encoding for evolving the placement,
  density, and connectivity of neurons</p></title>
  <aug>
    <au><snm>Risi</snm><fnm>S</fnm></au>
    <au><snm>Stanley</snm><fnm>KO</fnm></au>
  </aug>
  <source>Artificial life</source>
  <pubdate>2012</pubdate>
  <volume>18</volume>
  <issue>4</issue>
  <fpage>331</fpage>
  <lpage>363</lpage>
</bibl>

<bibl id="B12">
  <title><p>{Exploiting Open-Endedness to Solve Problems Through the Search for
  Novelty}</p></title>
  <aug>
    <au><snm>Lehman</snm><fnm>J</fnm></au>
    <au><snm>Stanley</snm><fnm>KO</fnm></au>
  </aug>
  <source>Artificial Life XI</source>
  <pubdate>2008</pubdate>
  <fpage>329</fpage>
  <lpage>-336</lpage>
</bibl>

<bibl id="B13">
  <title><p>Complex network measures of brain connectivity: Uses and
  interpretations</p></title>
  <aug>
    <au><snm>Rubinov</snm><fnm>M</fnm></au>
    <au><snm>Sporns</snm><fnm>O</fnm></au>
  </aug>
  <source>Neuroimage</source>
  <pubdate>2010</pubdate>
  <volume>52</volume>
  <issue>3</issue>
  <fpage>1059</fpage>
  <lpage>1069</lpage>
</bibl>

<bibl id="B14">
  <title><p>Multiobjective evolutionary algorithms: a comparative case study
  and the strength Pareto approach</p></title>
  <aug>
    <au><snm>Zitzler</snm><fnm>E.</fnm></au>
    <au><snm>Thiele</snm><fnm>L.</fnm></au>
  </aug>
  <source>IEEE Transactions on Evolutionary Computation</source>
  <pubdate>1999</pubdate>
  <volume>3</volume>
  <issue>4</issue>
  <fpage>257</fpage>
  <lpage>-271</lpage>
</bibl>

<bibl id="B15">
  <title><p>Steady state vs. generational genetic algorithms: A comparison of
  time complexity and convergence properties</p></title>
  <aug>
    <au><snm>Noever</snm><fnm>D</fnm></au>
    <au><snm>Baskaran</snm><fnm>S</fnm></au>
  </aug>
  <source>SFI Working Paper</source>
  <pubdate>1992</pubdate>
  <issue>1992-07-032</issue>
</bibl>

<bibl id="B16">
  <title><p>Fast generation of regular graphs and construction of
  cages</p></title>
  <aug>
    <au><snm>Meringer</snm><fnm>M</fnm></au>
  </aug>
  <source>Journal of Graph Theory</source>
  <pubdate>1999</pubdate>
  <volume>30</volume>
  <issue>2</issue>
  <fpage>137</fpage>
  <lpage>146</lpage>
</bibl>

<bibl id="B17">
  <title><p>Complex networks: Structure and dynamics</p></title>
  <aug>
    <au><snm>Boccaletti</snm><fnm>S.</fnm></au>
    <au><snm>Latora</snm><fnm>V.</fnm></au>
    <au><snm>Moreno</snm><fnm>Y.</fnm></au>
    <au><snm>Chavez</snm><fnm>M.</fnm></au>
    <au><snm>Hwang</snm><fnm>D. U.</fnm></au>
  </aug>
  <source>Physics reports</source>
  <pubdate>2006</pubdate>
  <volume>424</volume>
  <issue>4</issue>
  <fpage>175</fpage>
  <lpage>308</lpage>
</bibl>

<bibl id="B18">
  <title><p>Genetic search strategies in multicriterion optimal
  design</p></title>
  <aug>
    <au><snm>Hajela</snm><fnm>P.</fnm></au>
    <au><snm>Lin</snm><fnm>C. Y.</fnm></au>
  </aug>
  <source>Structural Optimization</source>
  <pubdate>1992</pubdate>
  <volume>4</volume>
  <issue>2</issue>
  <fpage>99</fpage>
  <lpage>107</lpage>
</bibl>

<bibl id="B19">
  <title><p>Multiple objective optimization with vector evaluated genetic
  algorithms</p></title>
  <aug>
    <au><snm>Schaffer</snm><fnm>J. D.</fnm></au>
  </aug>
  <source>Proceedings of the International Conference on Genetic Algorithm and
  their Applications</source>
  <editor>John J. Grefenstette</editor>
  <pubdate>1985</pubdate>
  <fpage>93</fpage>
  <lpage>100</lpage>
</bibl>

<bibl id="B20">
  <title><p>An evolutionary algorithm for constrained multi-objective
  optimization</p></title>
  <aug>
    <au><snm>Jimenez</snm><fnm>F.</fnm></au>
    <au><snm>Gomez Skarmeta</snm><fnm>A.F.</fnm></au>
    <au><snm>Sanchez</snm><fnm>G.</fnm></au>
    <au><snm>Deb</snm><fnm>K.</fnm></au>
  </aug>
  <source>Proceedings of the 2002 Congress on Evolutionary Computation</source>
  <publisher>IEEE</publisher>
  <pubdate>2002</pubdate>
  <volume>2</volume>
  <fpage>1133</fpage>
  <lpage>1138</lpage>
</bibl>

</refgrp>
} % end of \BMCxmlcomment
}

\end{bmcformat}
\end{document}